\documentclass[aps,prb,twocolumn]{revtex4-1}
\usepackage[english]{babel}
\usepackage{amsmath,amssymb,graphicx,hyperref}

\newcommand{\cdummy}{\cdot}
\newcommand{\mathd}{\mathrm{d}}
\newcommand{\nocomma}{}
\newcommand{\tmmathbf}[1]{\ensuremath{\boldsymbol{#1}}}
\newcommand{\tmop}[1]{\ensuremath{\operatorname{#1}}}

\begin{document}

\title{Straightforward computation of high-pressure elastic constants using
Hooke's law: A prototype of metal Ru}

\author{}
\noaffiliation

\author{Zhong-Li Liu}
\email{zl.liu@163.com}
\affiliation{ 
	College of Physics and Electric Information, Luoyang Normal University, Luoyang 471934, China
}%

\author{Xiu-Lu Zhang}

\affiliation{%
	Laboratory for Extreme Conditions Matter Properties, Southwest University of Science and Technology, 621010 Mianyang, China
}%

\author{Cuan-Cuan Zhu}
\affiliation{ 
	College of Physics and Electric Information, Luoyang Normal University, Luoyang 471934, China
}%
\affiliation{%
	School of Materials Science and Engineering, Henan Polytechnic University, Jiaozuo 454000, China
}%

\author{Hai-Yan Wang}
\email{wanghy@hpu.edu.cn}

\affiliation{%
School of Materials Science and Engineering, Henan Polytechnic University, Jiaozuo 454000, China
}%

\begin{abstract}
  In this paper, we did a systematic comparative study on the accuracy of two
  computational methods of elastic constants combined with the density
  functional theory (DFT), the stress-strain method and the energy-strain
  method. We took metal Ru as a prototype to compare its high-pressure elastic
  constants calculated by our present stress-strain method with the previous
  energy-strain results by others. Although the two methods yielded almost the
  same accuracy of high-pressure elastic constants for Ru, our stress-strain
  method directly based on the Hooke's law of elasticity theory is much
  straightforward and simple to implement. However, the energy-strain method
  needs complicated pressure corrections because of the pressure effects on
  the total energy. Various crystal systems have various pressure correction
  methods. Hence, the stress-strain method is preferred to calculate the
  high-pressure elastic constants of materials. Furthermore, we analyzed the
  variations of the elastic moduli, elastic anisotropy, sound velocities,
  Debye temperature of Ru with pressure.
\end{abstract}

{\maketitle}

\section{Introduction}

Elastic constants are key parameters for us to understand the physical and
even chemical properties of materials. They correlate closely with various
physical properties, such as mechanical (hardness) {\cite{Tehrani2019}},
thermodynamic {\cite{Golesorkhtabar2013}}, melting {\cite{Ashcroft1976}},
acoustic {\cite{Blanchfield1979}}, and so on. Elastic constants are frequently
used in many fields, including materials science, physics, geophysics,
chemical, condensed matter, and engineering and technology. Hence, the elastic
properties of materials continue attracting much attention from researchers.

There are generally two methods used for calculating the second-order elastic
constants of materials. One is the strain-energy method based on the
relationship between specific energy and strain magnitude
{\cite{Sinko2002,Sinko2005,Sinko2008,Perger2009,Wang2019,Golesorkhtabar2013}}.
The other is the stress-strain method {\cite{Golesorkhtabar2013,Yu2010,Zhang2018}}
fundamentally based on the Hooke's law of elasticity theory. The two methods
combined with the state-of-the-art density functional theory (DFT) make the
calculations of elastic constants simple. At zero pressure, the two methods
both easily yield identical results; however, at high pressure the
energy-strain method needs complex pressure corrections
{\cite{Sinko2002,Sinko2008}} to achieve effective elastic constants, with
different correction formulas for different crystal systems
{\cite{Sinko2008}}. While, at high pressure the stress-strain method is still
straightforward and simple without pressure corrections, directly according to
the Hooke's law like the case of zero pressure.

In this paper, we have a comparative study on the two methods in the
calculations of high-pressure elastic constants of materials. We here take
metal Ru as the prototype to compare the results of the two methods combined
with DFT. Since the high-pressure elastic constants of Ru has already been
calculated using the energy-strain method with pressure corrections by
Lugovskoy et al.{\cite{Lugovskoy2014}}. We only re-calculated its
high-pressure elastic constants using the stress-strain method. The details of
computational algorithm are introduced in Ref.{\cite{Liu2020}}, and our
recently developed ElasTool package {\cite{Liu2020-2}} was used to calculate
the elastic constants of Ru. We find that our stress-strain method yielded the
identical high-pressure elastic constants of Ru, compared to the previous
energy-strain method {\cite{Lugovskoy2014}}.

\section{Computational details}

The original Hooke's law states that the force ($F$) exerted on a spring is
linearly proportional to the spring's deformation magnitude ($x$)---that is,
$F = k \nocomma x$, where $k$ is a proportional coefficient reflecting its
stiffness, and $x$ is small compared to the possible maximum deformation of
the spring. The modern elasticity theory extends Hooke's law to state that the
strain (deformation) of a crystal is proportional to the stress applied to it.
However, since general stresses and strains are tensors, the proportional
coefficient is no longer \ just a single number, but rather a tensor
represented by a matrix.\label{method}

\subsection{ Elastic constants computation method}

According to Hooke's law, within the linear elastic regime of a crystal the
relation between the stresses $\sigma_i$ and its induced strains
$\varepsilon_j$ is,
\begin{equation}
  \sigma_i = \sum_{j = 1}^6 C_{i j} \varepsilon_j \label{hooke}
\end{equation}
where the $C_{i j}$ are the elastic constant of the crystal.

We can obtain all the elastic constants of a material from Eq.(\ref{hooke}),
by applying the strain $\varepsilon_j$ and calculating the corresponding
stresses, The deformation matrix applied is
\begin{equation}
  \tmmathbf{D}=\tmmathbf{I}+\tmmathbf{\varepsilon},
\end{equation}
where $\tmmathbf{I}$ is the $3 \times 3$ unit matrix, and
$\tmmathbf{\varepsilon}$ is the strain-matrix in Voigt notation. In the 3D
case, the strain matrix is
\begin{equation}
  \tmmathbf{\varepsilon}= \left[ \begin{array}{l}
    \varepsilon_1 \quad \frac{\varepsilon_6}{2} \quad
    \frac{\varepsilon_5}{2}\\
    \frac{\varepsilon_6}{2} \quad \varepsilon_2 \quad
    \frac{\varepsilon_4}{2}\\
    \frac{\varepsilon_5}{2} \quad \frac{\varepsilon_4}{2} \quad \varepsilon_3
  \end{array} \right] .
\end{equation}

The deformed crystal lattice vector is
\begin{equation}
  \tmmathbf{A}' =\tmmathbf{A} \cdummy \tmmathbf{D}
\end{equation}
where $\tmmathbf{A}$ is the crystal lattice vector without deformation.

\subsection{Elastic moduli and anisotropy}

From the elastic constants, the elastic moduli can be readily derived. The
Voigt and Reuss elastic moduli for different crystal systems are calculated
according to Ref. {\cite{Wu2007}}. According to the Voigt--Reuss--Hill
approximations {\cite{Hill1952}}, the arithmetic average of Voigt and Reuss
bounds is
\begin{equation}
  B = B_{\tmop{VRH}} = \frac{B_V + B_R}{2}, \label{bvrh}
\end{equation}
and
\begin{equation}
  G = G_{\tmop{VRH}} = \frac{G_V + G_R}{2} \label{gvrh}
\end{equation}
respectively.

Young's modulus $E$ is calculated by
\begin{equation}
  E = \frac{9 B \nocomma G}{3 B + G},
\end{equation}
and Poisson's ratio is
\begin{equation}
  \nu = \frac{3 B - 2 G}{2 (3 B + G)} . \label{nu}
\end{equation}
The elastic anisotropy is a direct reflect of spatial anisotropy of chemical
bonding. Chung and Buessem defined an elastic anisotropy index
{\cite{Chung1967}},
\begin{equation}
  A^C = \frac{G_V - G_R}{G_V + G_R} . \label{ac}
\end{equation}
Ranganathan and Ostoja-Starzewski proposed a more general anisotropy index
called the universal elastic anisotropy index {\cite{Ranganathan2008}},
\begin{equation}
  A^U = 5 \frac{G_V}{G_R} - \frac{B_V}{C_R} - 6 \geqslant 0. \label{au}
\end{equation}
$A^U = 0$ corresponds to the locally isotropic crystals \
{\cite{Ranganathan2008}}.

\subsection{The calculation details of stress tensors}

The crystal Ru has a hexagonal closely-packed (hcp) structure up to 600 GPa.
Before the calculations of elastic constants, the crystal structure of Ru was
first optimized within the framework of DFT at each pressure. Then the stress
components were calculated accurately after atomic positions were relaxed
under specific deformations applied according to our recently proposed ``the
optimized high efficiency strain-matrix sets (OHESS)''{\cite{Liu2020}}. The
relaxations stopped after the forces acted on each atom were less than 0.02
eV/{\r A}. The projector augmented wave (PAW) method {\cite{Blochl1994}} as
implemented in the Vienna ab initio simulation package (VASP)
{\cite{Kresse1999,Kresse1996}} was adopted in all the structural optimizations
and stress computations in this work. The Perdew, Becke and Ernzerhof (PBE)
{\cite{Perdew1997}} generalized gradient approximation (GGA) was used for the
exchange-correlation functional. The energy cutoff values were set to ensure
energy to be converged to $10^{- 6}$ eV.

\section{Results and discussions\quad}

\subsection{Elastic constants}

The calculated lattice parameters accords very well with both experimental and
others' theoretical results, as shown in Table \ref{lattpar}. Also listed in
Table \ref{lattpar} are our calculated elastic constants of Ru at 0 GPa. We
see that our elastic constants are in reasonably good with experimental data
and others' calculated results. Especially, we note that our stress-strain
results are in very good agreement with the results from the energy-strain
method by Lugovskoy et al.{\cite{Lugovskoy2014}}. It is not surprising because
they are the zero-pressure elastic constants, not considering pressure effects
in both the stress-strain and energy-strain methods.

\begin{table*}[htp!]
	  \caption{The comparison of our calculated high-pressure elastic constants of
		Ru with those from Ref.{\cite{Lugovskoy2014}}\label{parameters}. The atomic
		volume ($V$) is in {\r A}$^3$, and pressure ($P$) and elastic constants
		$C_{i \nocomma j}$ are in GPa.\label{ec-p}}
  \begin{tabular}{llllllllll}
  	\hline
  	\hline
    & $V_0$ ({\r A}$^3$) & $a$ ({\r A}) & $c / a$ & $C_{11}$ & $C_{12}$ &
    $C_{13}$ & $C_{33}$ & $C_{44}$ & \\
    \hline
    & 13.75 & 2.72 & 1.577 & 583.2 & 184.3 & 183.1 & 652.2 & 191.9 & This
    work\\
    & 13.76 & 2.72 & 1.578 & 577.4 & 176.7 & 170.9 & 644.2 & 190.4 &
    Calc.{\cite{Lugovskoy2014}}\\
    & 13.24 & 2.68 & 1.584 & 701.0 & 196.2 & 187.4 & 774.5 & 240.0 &
    Calc.{\cite{Fast1995}}\\
    &  &  &  & 627.9 & 154.2 & 125.5 & 565.4 & 150.0 &
    Calc.{\cite{Pandey2009}}\\
    & 13.51 & 2.70 & 1.585 &  &  &  &  &  & Expt.{\cite{Olijnyk2001}}\\
    & 13.49 & 2.71 & 1.552 &  &  &  &  &  & Expt.{\cite{Kittel1956}}\\
    & 14.48 & 2.71 & 1.582 & 576.3 & 187.2 & 167.3 & 640.5 & 189.1 &
    Expt.{\cite{Dirts2003}}\\
    $\mathd C_{i \nocomma j} / \mathd P$ &  &  &  & 7.20 & 3.47 & 3.12 & 7.78
    & 1.74 & This work\\
    &  &  &  & 7.16 & 3.26 & 3.24 & 7.65 & 1.69 &
    Calc.{\cite{Lugovskoy2014}}\\
    &  &  &  &  &  &  &  & 1.67 & Expt.{\cite{Olijnyk2001}}\\
    \hline
    \hline
  \end{tabular}
  \caption{The lattice parameters of Ru, in comparison with experimental data
  and others' theoretical values.\label{lattpar}}
\end{table*}

In order to compare the stress-strain method with the energy-strain method
(Ref. {\cite{Lugovskoy2014}}), we accurately tuned our pressures to correspond
to those calculated in Ref. {\cite{Lugovskoy2014}}. The maximum difference in
pressure is not larger than 0.3 GPa, as shown in Table \ref{ec-p}. The
corresponding mean atomic volumes are also in good agreement with each other.
The maximum of mean atomic volume differences are not larger than 0.04 {\r
A}$^3$, below 500 GPa. The high-pressure elastic constants are listed in Table
\ref{ec-p}, from which we can numerically compare the elastic constants from
the two different methods, the stress-strain and energy-strain methods. The
pressure derivative of elastic constants $\mathd C_{i \nocomma j} / \mathd P$
are shown in Table \ref{lattpar}. It is surprisingly good that our
stress-strain values agree with the energy-strain values that were calculated
with complex pressure corrections by Lugovskoy et al. {\cite{Lugovskoy2014}}.
This indicates that, although the stress-strain method is rather
straightforward to calculate the high-pressure elastic constants, it can reach
the same accuracy of the energy-strain method with complicated pressure
corrections.

\begin{table*}[htp!]
	  \caption{The comparison of our calculated high-pressure elastic constants of
		Ru with those from Ref.{\cite{Lugovskoy2014}}\label{parameters}. The atomic
		volume ($V$) is in {\r A}$^3$, and pressure ($P$) and elastic constants
		$C_{i \nocomma j}$ are in GPa.\label{ec-p}}
  \begin{tabular}{llllllll}
    \hline
    \hline
    & $V$ & $P$ & $C_{11}$ & $C_{12}$ & $C_{13}$ & $C_{33}$ & $C_{44}$\\
    \hline
    This work & 12.30 & 46.5 & 912.4 & 344.1 & 326.9 & 1009.7 & 271.7\\
    Ref.{\cite{Lugovskoy2014}} & 12.30 & 46.5 & 908.8 & 327.9 & 309.4 & 998.5
    & 268.5\\
    This work & 11.29 & 98.3 & 1232.6 & 509.3 & 471.7 & 1360.0 & 345.2\\
    Ref.{\cite{Lugovskoy2014}} & 11.31 & 98.5 & 1236.0 & 483.9 & 452.4 &
    1352.0 & 341.8\\
    This work & 10.92 & 124.7 & 1384.0 & 590.1 & 542.0 & 1527.5 & 379.0\\
    Ref.{\cite{Lugovskoy2014}} & 10.94 & 124.5 & 1391.0 & 558.9 & 521.2 &
    1516.0 & 374.9\\
    This work & 9.66 & 251.9 & 2046.2 & 965.0 & 862.7 & 2269.8 & 518.4\\
    Ref.{\cite{Lugovskoy2014}} & 9.69 & 251.9 & 2084.0 & 908.4 & 839.0 &
    2264.0 & 517.5\\
    This work & 8.50 & 454.2 & 2977.1 & 1526.2 & 1341.4 & 3340.3 & 699.5\\
    Ref.{\cite{Lugovskoy2014}} & 8.54 & 454.5 & 3063.0 & 1420.0 & 1326.0 &
    3346.0 & 707.7\\
    This work & 7.96 & 596.5 & 3592.9 & 1908.0 & 1661.8 & 4047.7 & 807.9\\
    Ref.{\cite{Lugovskoy2014}} & 8.10 & 596.5 & 3719.0 & 1772.0 & 1653.0 &
    4081.0 & 825.8\\
    \hline
    \hline
  \end{tabular}

\end{table*}

\begin{figure}[h]
  \resizebox{8cm}{6.4cm}{\includegraphics{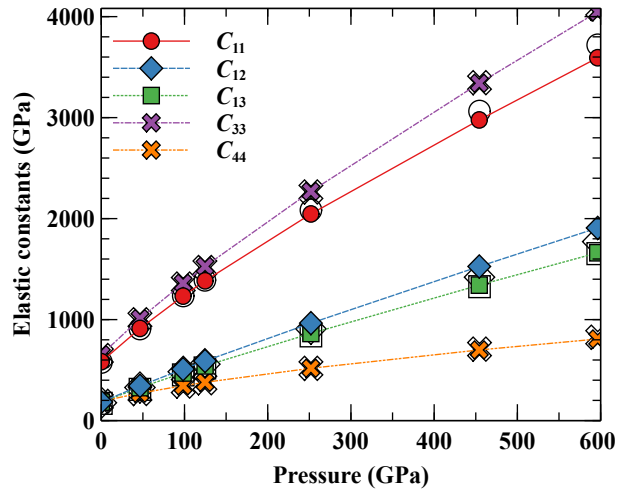}}
  \caption{The calculated elastic constants of Ru versus pressure (solid
  symbols), in comparison with the calculated results from
  Ref.{\onlinecite{Lugovskoy2014}} (open symbols).\label{fig-ec-p}}
\end{figure}

\begin{figure}[h]
  \resizebox{8cm}{6.4cm}{\includegraphics{{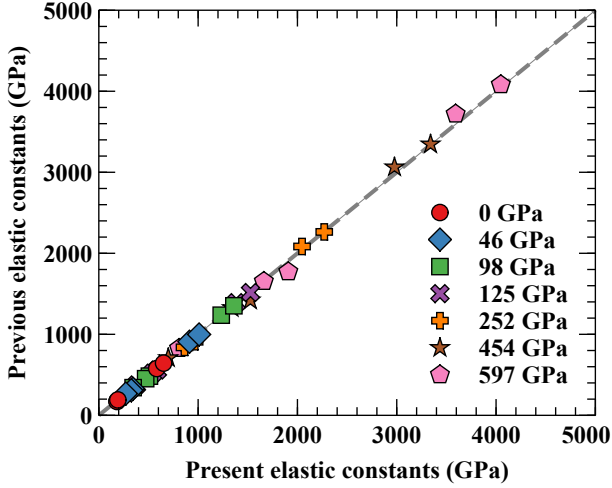}}}
  \caption{Comparison of the present calculated elastic constants and the
  previous results from Ref. {\onlinecite{Lugovskoy2014}}\label{fig2}.}
\end{figure}

To clearly compare the high-pressure elastic constants from the two methods,
we presented all the elastic constants versus pressure data in Fig.
\ref{fig-ec-p}. Except for the slight differences in $C_{11}$ and $C_{12}$
beyond 400 GPa, other elastic constants ($C_{13}, C_{33}$, and $C_{44}$) are
in excellent agreement with each other for the two methods, in the whole
pressure range from 0 up to $\sim$600 GPa. We also compare totally all the
high-pressure elastic constants from the two methods in Fig. \ref{fig2}, from
which we again find the satisfactory agreements between the stress-strain and
energy-strain methods. In order to statistically evaluate the agreement of our
results with the previous ones, we calculated the coefficient of determination
$R^2$ score {\cite{R2}} of present compared to previous results. The evaluated
$R^2 = 0.9982$, implying an excellent agreement quantitatively.

\subsection{Elastic moduli and elastic anisotropy}

From our high-pressure elastic constants, we calculated the bulk modulus $B$,
shear modulus $G$, and Young's modulus $E$ according to Eqs. (\ref{bvrh}),
(\ref{gvrh}), and (\ref{nu}), respectively, and show them in Table.
\ref{moduli}. These three elastic moduli all increase monotonously with
pressure, implying the rigidity of Ru increases with pressure. We also
calculated the Poisson's ratio of Ru and found that it increases from 0.2431
to 0.3367 at pressure ranging from 0 to $\sim$600 GPa.

In order to understand the variation of metallic bonding in Ru with pressure,
we analyzed its elastic anisotropy properties. First, according to Eqs.
(\ref{ac}) and (\ref{au}), we calculated the elastic anisotropy indexes which
can overall reflect the bonding characteristics of materials. As shown in
Table \ref{moduli}, both $A^U$ and $A^C$ have very small variations at
pressures from 0 up to $\sim$600 GPa. This indicates that the bonding
properties keeps almost invariant in the hcp-Ru.

\begin{figure}[h]
  \resizebox{4cm}{!}{\includegraphics{{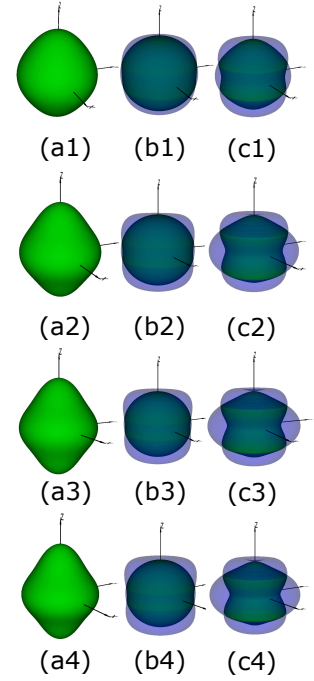}}}
  \caption{The spatial variations of the (a) Young's modulus, (b) shear
  modulus, and (c) Poisson's ratio of metal Ru at pressures (1) 0.03 GPa, (2)
  251.9 GPa, (3) 454.2 GPa, and (4) 596.5 GPa, respectively.\label{spacial}}
\end{figure}

Second, we plotted the spatial variations of the Young's modulus, shear
modulus, and Poisson's ratio at different pressures in Fig. \ref{spacial}. It
can be clearly seen that the spatial elastic anisotropy of Ru almost remains
unchanged with pressure. This accords well with the conclusion drawn from the
elastic anisotropy indexes analysis.

\begin{table}[htp!]
	  \caption{The high-pressure elastic moduli, Poisson's ratio, and elastic
		anisotropy indexes of Ru. The moduli are all in GPa.\label{moduli}}
  \begin{tabular}{lllllll}
    \hline
    \hline
    $P$ & $B$ & $G$ & $E$ & $\nu$ & $A^U$ & $A^C$\\
    \hline
    0.0 & 324. & 200.8 & 499.3 & 0.2431 & 0.0200 & 0.0018\\
    46.5 & 536.3 & 287.2 & 731.2 & 0.2728 & 0.0266 & 0.0025\\
    98.3 & 747.5 & 367.3 & 946.9 & 0.2889 & 0.0343 & 0.0033\\
    124.7 & 849.0 & 404.3 & 1046.6 & 0.2945 & 0.0380 & 0.0037\\
    251.9 & 1304.4 & 557.3 & 1463.4 & 0.3130 & 0.0553 & 0.0054\\
    349.9 & 1632.0 & 658.1 & 1740.3 & 0.3223 & 0.0669 & 0.0066\\
    454.2 & 1967.4 & 758.0 & 2015.2 & 0.3293 & 0.0792 & 0.0078\\
    596.5 & 2410.0 & 883.4 & 2361.7 & 0.3367 & 0.0958 & 0.0094\\
    \hline
    \hline
  \end{tabular}

\end{table}

\subsection{Sound velocity and Debye temperature}

The phase velocity $v$ and polarization of the three waves along a fixed
propagation direction defined by the unit vector $n_i$ are given by Cristoffel
equation,
\begin{equation}
  (C_{i \nocomma j \nocomma k \nocomma l} n_j n_k - \rho v^2 \delta_{i
  \nocomma j}) u_i = 0,
\end{equation}
where $C_{i \nocomma j \nocomma k \nocomma l}$ is the
fourth-rank tensor description of the elastic constants, $n$ is the
propagation direction, and $u$ the polarization vector.
\begin{figure}[htp!]
	\resizebox{8cm}{6.4cm}{\includegraphics{{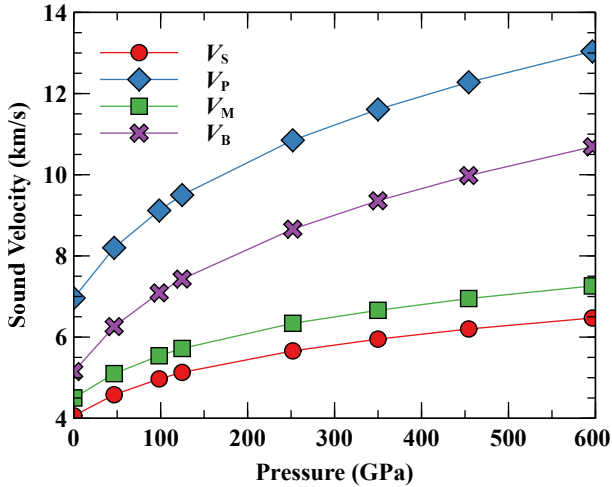}}}
	\caption{The sound velocities $V_B$, $V_P$, $V_S$, and $V_M$ of metal Ru as
		functions of pressure.\label{velocity}}
\end{figure}
\begin{table}[htp!]
	\caption{The Debye temperatures of Ru at different pressure.\label{debye}}
	\begin{tabular}{ll}
		\hline
		\hline
		$P$ (GPa) & $\Theta_D$ (K)\\
		\hline
		0.0 & 444\\
		46.5 & 522\\
		98.3 & 584\\
		124.7 & 609\\
		251.9 & 703\\
		349.9 & 756\\
		454.2 & 804\\
		596.5 & 859\\
		\hline
		\hline
	\end{tabular}
	
\end{table}

The longitudinal wave velocity is
\begin{equation}
  v_P = \sqrt{\frac{B + 4 G / 3}{\rho}}, \label{vp}
\end{equation}
and the body wave velocity is
\begin{equation}
  v_B = \sqrt{\frac{B}{\rho}} .
\end{equation}
The shear wave velocity is calculated by
\begin{equation}
  v_S = \sqrt{\frac{G}{\rho}} .
\end{equation}
From $v_S$ and $v_P$, we can obtain the mean wave velocity via
\begin{equation}
  v_m = \left[ \frac{1}{3} \left( \frac{2}{v_S^3} + \frac{1}{v_P^3} \right)
  \right]^{- 1 / 3} . \label{vm}
\end{equation}
From the mean wave velocity data, we can calculate the Debye temperature from
\begin{equation}
  \Theta_D = \frac{h}{k} \left[ \frac{3 n}{4 \pi} \left( \frac{N_A \rho}{M}
  \right) \right]^{1 / 3} v_m \label{td}
\end{equation}
where $h$ is Planck's constant, $k$ the Boltzmann's constant, $N_A$ the
Avogadro's number, $n$ the number of atoms in the unit cell, $M$ the weight of
the unit cell, and $\rho$ the density.

We show the calculated sound velocities in Fig. \ref{velocity}, from which we
see the monotonous increases in all these sound velocities. This indirectly
reflects the strengthening of the metallic bondings in Ru with increasing
pressure.

Debye temperature $\Theta_D$ corresponds to the temperature of a crystal's
highest normal mode of lattice vibration, and it closely correlates the
elastic properties with the thermodynamic properties {\cite{Luo2008}}. The
calculated Debye temperatures are listed in Table \ref{debye}. The increasing
of $\Theta_D$ with pressure again reflects the enhancing of the metallic
bonding in Ru.

\section{Conclusions}

In conclusion, we systematically compared the elastic constants of metal Ru
calculated by the stress-strain method with those previously obtained by the
energy-strain method. Both the zero-pressure and high-pressure elastic
constants of Ru agree very well with each other for the two methods. But, our
stress-strain method is much straightforward to implement to achieve
high-pressure elastic constants. The energy-strain method depends on the
complex pressure corrections for obtaining high-pressure elastic constants of
materials, though the two methods can both reach the same accuracy. Thus the
stress-strain method is more preferred for calculating the high-pressure
elastic constants of materials. From the calculated elastic constants of Ru,
we also analyzed the variations of its high-pressure elastic moduli, Poisson's
ratio, elastic anisotropy, and Debye temperature with increasing pressure. All
these physical parameters increases monotonously with pressure, implying the
enhancement of rigidity of Ru under pressure. While, the elastic anisotropy
varies slightly with pressure, reflecting the slight spatial variations of
metallic bonding with pressure.

\section{Acknowledgments}

We acknowledge the supports from the National Natural Science Foundation of
China (41574076), the Key Research Scheme of Henan Universities (18A140024),
the Key Laboratory of the Electromagnetic Transformation and Detection of
Henan province, and the Research Scheme of LYNU Innovative Team under Grant
No. B20141679.

%
\end{document}